\documentclass[aps,pra,reprint,superscriptaddress]{revtex4-2}
\usepackage[utf8]{inputenc}
\usepackage{physics}
\usepackage{graphicx}
\usepackage{multirow}
\usepackage{hyperref}
\hypersetup{colorlinks=true, citecolor=red}
\usepackage{cleveref}
\usepackage{xcolor}
\usepackage{bbold} 
\usepackage{tabularx} 
\usepackage{soul}  
\usepackage{tikz}
\usetikzlibrary{quantikz2}

\usepackage[caption = false]{subfig}

\definecolor{pink}{HTML}{F282B4}

\newcommand{\KL}{D_{\text{KL}}}

\newcommand{\Wh}{\mathcal{W}_{\mathrm{h}}}

\newcommand{\sqRBM}{\mathrm{sqRBM}}

\newcommand{\id}{\mathbb{1}}
\newcommand{\X}{\mathbb{X}}
\newcommand{\Y}{\mathbb{Y}}
\newcommand{\Z}{\mathbb{Z}}

\newcommand{\sZ}{\sigma^{\scriptstyle \mathrm{Z}}}

\newcommand{\aZ}{a^{\scriptstyle \mathrm{Z}}}

\newcommand{\Iv}{\id^{\otimes n}}
\newcommand{\Ih}{\id^{\otimes m}}

\usepackage{makecell} 
\usepackage{mathtools}
\usepackage{tabularx} 
\usepackage{amsmath}
\usepackage{amssymb}
\usepackage{amsthm}
\usepackage{braket}
\usepackage{algorithm}
\usepackage{algorithmic}

\newcolumntype{C}[1]{>{\centering\arraybackslash}p{#1}}

\newtheorem{theorem}{Theorem}
\newtheorem{definition}{Definition}

\begin{document}

\title{Sample-based training of quantum generative models}

\date{\today}

\author{Maria Demidik}
\email{maria.demidik@desy.de}
\affiliation{Deutsches Elektronen-Synchrotron DESY, Platanenallee 6, 15738
Zeuthen, Germany}
\affiliation{Computation-Based Science and Technology Research Center, The Cyprus Institute, 20 Kavafi Street, 2121 Nicosia, Cyprus}

\author{Cenk T\"uys\"uz}
\affiliation{European Organisation for Nuclear Research (CERN), Espl. des Particules 1211 Geneva 23, Switzerland}

\author{Michele Grossi}
\affiliation{European Organisation for Nuclear Research (CERN), Espl. des Particules 1211 Geneva 23, Switzerland}

\author{Karl Jansen}
\affiliation{Computation-Based Science and Technology Research Center, The Cyprus Institute, 20 Kavafi Street, 2121 Nicosia, Cyprus}
\affiliation{Deutsches Elektronen-Synchrotron DESY, Platanenallee 6, 15738
Zeuthen, Germany}

\begin{abstract} 
Quantum computers can efficiently sample from probability distributions that are believed to be classically intractable, providing a foundation for quantum generative modeling. However, practical training of such models remains challenging, as gradient evaluation via the parameter-shift rule scales linearly with the number of parameters and requires repeated expectation-value estimation under finite-shot noise. We introduce a training framework that extends the principle of contrastive divergence to quantum models. By deriving the circuit structure and providing a general recipe for constructing it, we obtain quantum circuits that generate the samples required for parameter updates, yielding constant scaling with respect to the cost of a forward pass, analogous to backpropagation in classical neural networks. Numerical results demonstrate that it attains comparable accuracy to likelihood-based optimization while requiring substantially fewer samples. The framework thereby establishes a scalable route to training expressive quantum generative models directly on quantum hardware.
\end{abstract}

\maketitle

\section{Introduction}\label{sec:intro}

Foundational results in quantum computational complexity provide strong evidence that quantum devices can efficiently sample from distributions that are believed to be intractable for classical computers, unless the polynomial hierarchy collapses. Prominent examples include sampling from instantaneous quantum polynomial-time (IQP) circuits~\cite{Bremner_2010}, random circuit sampling~\cite{arute_quantum_2019}, and boson sampling~\cite{10.1145/1993636.1993682}. While the hardness of these tasks relies on specific complexity-theoretic assumptions such as anti-concentration and average-case hardness, they establish credible theoretical separations between classical and quantum sampling tasks. These separations have motivated the development of quantum generative models that aim to translate this sampling advantage into practical machine learning settings~\cite{huang2025generative}. This line of reasoning has inspired a growing body of work in quantum machine learning (QML), where quantum circuits are explored as generative models capable of representing and sampling from complex probability distributions.

Beyond such demonstrations of sampling capability, the development of scalable and trainable quantum generative models remains a major challenge. In particular, achieving efficient and reliable training of QML models has proven difficult, primarily due to the challenges associated with accurate gradient estimation. Most existing QML approaches employ variational circuits~\cite{tuysuz2025} whose gradients are computed using the parameter-shift rule or finite-difference estimators~\cite{PhysRevLett.118.150503,PhysRevA.98.032309,PhysRevA.99.032331,Banchi2021measuringanalytic}. Both methods require separate circuit executions for each variational parameter, leading to a computational cost that scales linearly with the number of parameters~\cite{Wierichs2022generalparameter}. This cost is further amplified by finite-shot noise, which introduces stochastic fluctuations in gradient estimates and necessitates repeated measurements to achieve sufficient precision~\cite{Kreplin2024reductionoffinite}. Moreover, many QML models are affected by barren plateaus, where gradients vanish exponentially with system size~\cite{mcclean2018}. As a result, the number of circuit evaluations required to obtain statistically significant gradients also grows exponentially, severely limiting the scalability of variational training.

Unlike classical deep neural networks, QML models lack a true backpropagation-like mechanism that allows gradients to be efficiently propagated through circuit layers~\cite{NEURIPS2023_8c3caae2}. Several recent studies have explored quantum analogues of backpropagation. Ref.~\cite{NEURIPS2023_8c3caae2} proposed a hybrid approach that combines classical automatic differentiation with quantum gradient evaluation to partially reconstruct a backpropagation-like workflow, but the computational cost still scales unfavorably with circuit depth. An alternative strategy~\cite{Bowles2025backpropagation} has shown that favorable scaling can be achieved for specific structured-circuit architectures, for example by leveraging commuting gates, though at the cost of reduced model expressivity. Despite these advances, a general and scalable algorithm for gradient-based training of quantum models has yet to be established. Thus, the absence of efficient training strategies, rather than limitations in the expressive power of quantum models, remains the main bottleneck to realizing practical quantum generative learning.

One prominent alternative to backpropagation in classical machine learning is contrastive divergence (CD) learning~\cite{bm_cd}, a method widely used for training energy-based models~\cite{salakhutdinov_efficient_2012, salakhutdinov_deep_2009, du_improved_2021}. CD estimates gradients using samples from the data and model distributions, allowing all parameters to be updated simultaneously based on these samples. This procedure yields an asymptotically constant computational cost per training step, scaling with the cost of a single forward pass and remaining independent of the number of model parameters. In practice, CD is primarily employed to approximate likelihood gradients using short Gibbs sampling chains in restricted Boltzmann machines (RBM), thereby alleviating the need for costly sampling from the model distribution~\cite{bm_cd}. 

Quantum Boltzmann machines (QBM) extend classical Boltzmann machines to the quantum domain by replacing the classical energy function with a quantum Hamiltonian whose Gibbs state defines the model distribution~\cite{amin_quantum_2018}. This formulation allows QBMs to represent probability distributions that capture quantum correlations and interference effects, potentially increasing their expressivity beyond that of classical RBMs~\cite{tysz2024learning}. However, as with other QML models, training QBMs remains challenging due to the lack of backpropagation-like scaling in gradient computation~\cite{amin_quantum_2018, kieferova_tomography_2017, kappen_learning_2020, wiebe_generative_2019, coopmans_sample_2024, zoufal_variational_2021, demidik2025expressive, patel2024natural, kimura2025em}.

Building on these ideas, we propose a generalized CD algorithm for quantum models that achieves $\mathcal{O}(1)$ scaling in the cost of a forward pass. This scaling is analogous to the constant-cost updates of classical backpropagation. The generalized CD framework can be applied to QBMs and enables training of a quantum model entirely on quantum hardware.

Beyond providing a concrete solution for training quantum models, the generalized CD framework also addresses key limitations in obtaining gradients from data in quantum generative modeling. Most existing approaches require explicit access to the model probability distribution, and a widely adopted workaround is the use of a maximum mean discrepancy (MMD) loss~\cite{rudolph2024, recio-armengol_train_2025, kurkin2025universality}. While MMD enables gradient computation through a kernel defined on data samples, MMD-based training often proves restrictive and performs poorly for complex probability distributions~\cite{rudolph2024, herrerogonzalez2025born}. In particular, for QBMs, no approach has so far been able to train a model using data samples alone. Our method resolves this limitation, as CD computes gradients directly from data samples without requiring explicit access to the model distribution. 
 
In this work, we develop a generalized CD training strategy and showcase it using QBMs, demonstrating that it enables efficient learning from data samples without requiring explicit access to model probabilities while maintaining model expressivity. The remainder of this work is organized as follows. Section~\ref{sec:background} introduces the necessary notation, formally defines the QBM, and summarizes existing training strategies from the literature. In Section~\ref{sec:results}, we generalize CD to the QBM framework, present its corresponding quantum circuit construction, and define the complete training procedure. We then numerically validate the algorithm by comparing it against negative log-likelihood training, which requires explicit access to the data distribution. Our results show that the proposed method achieves comparable accuracy in reproducing the target data distribution while requiring significantly fewer total samples for training. The numerical trends, together with the underlying scaling arguments, indicate that this separation is expected to widen with increasing system size. Finally, Section~\ref{sec:discussion} provides a brief discussion of the limitations, implications, and future directions of this work.

\section{Background}\label{sec:background}

\subsection{Boltzmann machines}

A Boltzmann machine (BM) is a machine learning model inspired by statistical physics, particularly by systems such as the Ising model. It consists of binary visible and hidden units, each of which updates its state according to the states of its neighbors. Let $n$ denote the number of visible units and $m$ the number of hidden units. The visible units represent the observed data, while the hidden units capture higher-order correlations among them. The representational capacity of a BM increases with the number of hidden units, which determine the complexity of the correlations the model can learn~\cite{demidik2025expressive, le_roux_representational_2008}.

A joint probability distribution of visible and hidden units follows the Gibbs-Boltzmann distribution. Thus, at thermal equilibrium, the system is described by the Gibbs state given as
\begin{equation*}
    \rho = e^{- \beta H}/\mathcal{Z},\quad \mathcal{Z} = \Tr[e^{- \beta H}],  
\end{equation*}
where $H$ is a parameterized Hamiltonian encoding the system’s connectivity and interactions, $\beta = 1 / kT$ is the inverse temperature with $k$ denoting the Boltzmann constant and $T$ the temperature, and $\mathcal{Z}$ is the partition function summing over all possible configurations of the system. In this work, we define the Hamiltonian using Pauli operators $\{\id, \X, \Y, \Z\}$. 

In order to obtain the probability distribution over the visible configurations $p(v)$, one marginalizes over the hidden units. In the quantum framework, this marginalization corresponds to performing a projective measurement on the visible subsystem while tracing out the hidden subsystem:
\begin{equation}
p(v) = \Tr[(\Pi_v \otimes \Ih) \, \rho], \quad \Pi_v = \ket{v}\!\bra{v},
\label{eq:marginal-prob-visible}
\end{equation}
where $v \in \{0,1\}^n$ and $\Pi_v$ is the projection operator in the computational basis.

The representational capacity and computational tractability of BMs are largely determined by the connectivity structure between units~\cite{montufar_expressive_2014}. Although fully connected BMs provide a powerful and expressive framework for modeling complex probability distributions, they render both inference and training computationally demanding. To make learning tractable, it is common in the machine learning literature to impose a bipartite structure in which connections exist only between visible and hidden units, while lateral connections within each layer are absent. A BM with this restricted connectivity is known as a restricted Boltzmann machine (RBM)~\cite{smolensky_bm}. This restriction leads to conditional independence between units within the same layer, allowing efficient Gibbs sampling and gradient-based training that we discuss in the following subsections.

Classical BMs, including RBMs, are characterized by commuting Hamiltonians that correspond to stochastic Ising models.
Extending this formalism to non-commuting Hamiltonians gives rise to quantum Boltzmann machines (QBMs). The semi-quantum restricted Boltzmann machine (sqRBM) further unifies these two settings, providing a general framework that encompasses both classical and quantum RBMs for learning from classical data. For completeness, we restate the definition of sqRBMs as introduced in~\cite{demidik2025expressive}.

\begin{definition}[Semi-quantum RBM~\cite{demidik2025expressive}]
\label{def:sqrbm}
A semi-quantum restricted Boltzmann machine with $n$ visible and $m$ hidden units, denoted $\sqRBM_{n,m}$, is described by a parameterized Hamiltonian as follows:
\begin{equation}
H=\sum_{i=1}^n \aZ_i \sZ_i
  + \sum_{P\in\Wh}^{\mid \Wh \mid}\Big(
      \sum_{j=1}^m b_j^P\,\sigma_{n+j}^P
      + \sum_{i=1}^n\sum_{j=1}^m w_{ij}^P\,\sZ_i \sigma_{n+j}^P
    \Big),
\label{eq:sqrbm-hamiltonian}
\end{equation}
where $\boldsymbol{a} \in \mathbb{R}^{n}$, $\boldsymbol{b} \in \mathbb{R}^{|\Wh| \cdot m}$ and $\boldsymbol{w} \in \mathbb{R}^{|\Wh| \cdot n \cdot m}$ are the parameter vectors of the model. $\Wh$ is a non-commuting set of one-qubit Pauli operators that non-trivially act only on the hidden units. $\sigma^P_{i}$ is a length $n+m$ Pauli string acting non-trivially with Pauli operator $P$ on $i$-th qubit: $\sigma^P_{i} = \id^{\otimes (i-1)} \otimes P \otimes \id^{\otimes (n+m-i)}$.
\end{definition}

Alternatively, the Hamiltonian in Eq.~\eqref{eq:sqrbm-hamiltonian} can be expressed in the compact form as $H=\sum_i \theta_i H_i$, where $\boldsymbol{\theta}=\{\boldsymbol{a}, \boldsymbol{b}, \boldsymbol{w}\}$ denotes the set of parameters and $H_i$ the associated Pauli strings. We adopt this notation whenever it is convenient.

Choosing the restricted set $\Wh=\{\Z\}$ in Definition~\ref{def:sqrbm} recovers the classical RBM, in which all terms of the Hamiltonian commute, i.e., $\forall i:\,[H_i,H]=0$. By contrast, allowing the full set $\Wh=\{\X,\Y,\Z\}$ introduces non-commuting interactions, which define the sqRBM. Such sqRBMs achieve learning performance comparable to RBMs while requiring three times fewer hidden units~\cite{demidik2025expressive}. The reduction in the number of required hidden units for sqRBMs highlights their potential as early quantum generative models, particularly in settings with limited quantum resources. Unless stated otherwise, we adopt the convention that sqRBMs are defined using the full operator set \(\Wh=\{\X,\Y,\Z\}\) throughout the manuscript.

\subsection{Training Boltzmann machines}
\label{sec:quantum-training}

Given a Gibbs state, the parameters of both classical and quantum  Boltzmann machines are optimized such that the model reproduces the statistics of a target data distribution. This can be formalized as the minimization of the negative log-likelihood
\begin{equation*}
    \mathcal{L} = -\sum_v q(v) \log p(v),
\end{equation*}
where $q(v)$ denotes the empirical (target) distribution over visible configurations $v$, and $p(v)$ is the model distribution. Both distributions are normalized such that $\sum_v q(v) = \sum_v p(v) = 1$. 

It is important to note that, even if one could prepare the Gibbs state efficiently, evaluating gradients of $\mathcal{L}$ for general QBMs remains computationally intractable~\cite{amin_quantum_2018}, reflecting the intrinsic difficulty of gradient-based training in these models. Recent work has reduced the cost of gradient evaluation for sqRBMs by exploiting their structure~\cite{demidik2025expressive}. In what follows we focus on restricted architectures and restate the gradient expression for sqRBMs that is applicable to RBMs as well.

\begin{definition}[Gradients of $\sqRBM$~\cite{demidik2025expressive}]
\label{def:sqbm-grads}
Let $H$ denote the Hamiltonian of an $\sqRBM_{n,m}$, decomposed as $H = \sum_i \theta_i H_i$, where each $\theta_i$ is a real-valued model parameter. Let $\rho$ be the corresponding Gibbs state at inverse temperature $\beta = 1$, and let $\Pi_v$ denote the projection operator defined in Eq.~\eqref{eq:marginal-prob-visible}. Then, the gradient of the negative log-likelihood $\mathcal{L}$ with respect to  $\theta_i$ is given by
\begin{equation}
\partial_{\theta_i}\mathcal{L} = - \sum_v q(v)\,\Bigl( \Tr[\rho\, H_i] - \Tr[(\Pi_v \otimes \Ih)\,\rho\, H_i] \Bigr),
\label{eq:sqRBM-grad}
\end{equation}
where $q(v)$ is the target probability distribution.
\end{definition}

The expression in Eq.~\eqref{eq:sqRBM-grad} can be evaluated by computing the expectation values of the Hamiltonian terms $H_i$ over the Gibbs state $\rho$. In practice, however, this approach does not scale favorably and introduces several conceptual and computational challenges. 

First, the gradient rule assumes explicit access to the target probabilities $q(v)$, which is often unrealistic, as in many cases one only has access to a finite set of training samples drawn from the underlying data-generating process \footnote{For certain applications, such as fitting a distribution or using a Boltzmann machine as a sampler, this limitation may be less severe.}.  

Second, evaluating expectation values on projected Gibbs states introduces an additional layer of complexity. Specifically, one must prepare a distinct Gibbs state for each element in the non-zero support of the target distribution, $\mathrm{supp}(q) = \{v \mid q(v) > 0\}$. Consequently, the cost of gradient evaluation scales with $|\mathrm{supp}(q)|$ in the visible configuration space. Since many practically relevant distributions exhibit polynomial support, this requirement quickly becomes prohibitive: each training iteration may require the preparation of $\mathcal{O}(\mathrm{poly}(n))$ Gibbs states.  

Finally, each expectation value must be estimated to a target precision~$\epsilon$, which introduces an additional overhead of $\mathcal{O}(\epsilon^{-2})$ measurement shots due to statistical fluctuations. Taken together, the total cost of gradient evaluation thus scales as $\mathcal{O}\!\left(|\mathrm{supp}(q)| \epsilon^{-2}\right)$, which highlights the combined dependence on the support size of the target distribution and the desired estimation accuracy.  

While the scaling discussed above may seem unfavorable, it is important to note that RBMs and sqRBMs are not expected to suffer from barren plateaus~\cite{demidik2025expressive}, where the number of measurements required to estimate gradients grows exponentially with system size. Their bipartite connectivity and the local structure of their Hamiltonians prevent gradients from vanishing, and as a result, gradient magnitudes are expected to remain polynomially bounded. Nevertheless, the overall computational cost still depends on the support size and the desired precision, which introduces a non-negligible overhead. This highlights the practical importance of developing approximate or data-driven training methods that can alleviate these scaling constraints.

\subsection{Contrastive divergence learning}

In classical generative modeling, an efficient approach to learning complex high-dimensional distributions is to combine simpler probabilistic learning experts, as in mixture or product models~\cite{mclachlan_finite_2000, bm_cd}. These architectures support sample-based training procedures in which parameters are updated using samples drawn from both the data and the model distributions. RBMs belong to the class of product models, where stochastic gradient estimates can be obtained through contrastive divergence (CD) learning. We provide a formal description of CD training procedure below.

A compact expression of the parameter update rule~\cite{bm_cd, carreira-perpinan_contrastive_2005} can be obtained by rewriting the Definition~\ref{def:sqbm-grads} as
\begin{equation*}
    \partial_{\theta_i} \mathcal{L} =  \langle H_i \rangle_{\mathrm{data}} - \langle H_i \rangle_{\mathrm{model}} \, ,
\end{equation*}
where $\langle \cdot \rangle_{\mathrm{data}}$ denotes the expectation value evaluated on configurations of the empirical data and corresponds to the term with projective measurement in Eq.~\eqref{eq:sqRBM-grad}, and $\langle \cdot \rangle_{\mathrm{model}}$ denotes the corresponding expectation under the model.

Estimating expectations during training procedure then amounts to sampling from the corresponding Gibbs state. For general architectures this is classically intractable, since preparing exact Gibbs samples scales exponentially with the number of units~\cite{Ackley1985ALA, bm_hard_simulate}. A standard approximation is Gibbs sampling~\cite{Hinton1983OPTIMALPI}, where Markov chains are used to generate samples both conditioned on the data and from the model distribution. However, running long chains to equilibrium in high-dimensional state spaces is computationally expensive, particularly for model's expectations, and often yields high-variance gradient estimates~\cite{bengio_justifying_2009}.

In the case of RBM training, the absence of intra-layer connectivity allows visible and hidden units to be updated in parallel, making Gibbs sampling more tractable. Nevertheless, equilibrium sampling remains inefficient if long chains are required. To address this, CD was introduced in Ref.~\cite{bm_cd}, which replaces the need for equilibrium sampling with short chains. Instead of matching the data distribution directly to the model, CD compares the data with reconstructions obtained after $k$ Gibbs updates, yielding the objective
\begin{equation*}
    \mathrm{CD}_k = \KL (q \,\|\, p_{\mathrm{model}}) - \KL (p^k \,\|\, p_{\mathrm{model}}),
\end{equation*}
where $\KL$ is the Kullback-Leibler (KL) divergence, $q$ is the data distribution, $p_{\mathrm{model}}$ is the equilibrium distribution and $p^k$ denotes the distribution after $k$ sampling steps. In practice, small $k$ (often $k=1$) is sufficient to approximate gradient directions~\cite{carreira-perpinan_contrastive_2005, sutskever_convergence_2010}. This approximation introduces a bias but dramatically reduces computational costs, yielding gradients of the form~\cite{bm_cd}
\begin{equation*}
    \Delta \theta_i \propto \langle H_i \rangle_{\mathrm{data}} - \langle H_i \rangle_{p^k} \, . 
\end{equation*}
We provide a more detailed description of the CD algorithm in Appendix~\ref{app:cd}.

The applicability of CD hinges on the bipartite structure of RBMs, which ensures that the conditional distributions factorize:
\begin{align}
 p(h\,\mid\,v) &= \prod_{j=1}^m p(h_j\,\mid\,v), \quad p(v\,\mid\,h) = \prod_{i=1}^n p(v_i\,\mid\,h) \, ,
\end{align}
where $v_i$ and $h_j$ are the states of visible units $i$ and hidden unit $j$ correspondingly, $v \in \{0,1\}^n \, , h \in \{ 0,1 \}^m$.
These conditionals can be sampled exactly and in parallel, which is not the case for fully connected Boltzmann machines. As a result, CD has become the standard method for training RBMs in the classical machine learning literature. 

Although CD is computationally efficient for training RBMs and has been successfully applied in many domains~\cite{chen_continuous_2002, 10.5555/945365.964304, xuming_he_multiscale_2004}, it provides only a biased approximation of the true gradient~\cite{10.5555/2976040.2976240}. Ref.~\cite{sutskever_convergence_2010} demonstrated that the parameter updates obtained from CD do not follow the gradient of any objective function, raising theoretical concerns about the convergence of the algorithm. These challenges have motivated the development of alternative methods such as persistent contrastive divergence~\cite{younes_convergence_1999,Tieleman2008TrainingRB} or tensor network based training~\cite{carleo_solving_2017, chen_equivalence_2018, glasser_neural-network_2018, li_boltzmann_2021}. Despite that, CD remains the primary heuristic to approximate training of RBMs. 

\section{Results}\label{sec:results}

The close architectural correspondence between RBMs and sqRBMs suggests that CD learning can be extended to the quantum setting. Such an extension enables approximate, sample-based training of QBMs, mitigating the limitations associated with evaluating their gradients. The following subsections introduce the generalized CD framework.

\subsection{Conditional probabilities and circuit construction}

The evaluation of conditional probabilities is central to the training of BM models. In the case of sqRBMs, we establish that these probabilities can be expressed in a simple and exact form, enabled by the commutativity of the Hamiltonian with Pauli basis projectors. This provides a characterization of conditional probabilities that naturally translates into a circuit-level implementation. We express this result in the following theorem.

\begin{theorem}[Conditional probabilities of sqRBMs, informal]\label{theorem:cond-probs}
Let $H$ be the Hamiltonian of an sqRBM acting on $\mathcal H_{\mathrm{vis}} \otimes \mathcal H_{\mathrm{hid}}$ with $n$ visible and $m$ hidden qubits, and let $\beta>0$. Denote by $\Iv$ and $\Ih$ the identity operators on $\mathcal H_{\mathrm{vis}}$ and $\mathcal H_{\mathrm{hid}}$. For $v \in \{0,1\}^n$, let $\Pi_v^Z := \ket{v}\!\bra{v}$ be the projector onto the visible computational basis state. For the hidden subsystem, fix a Pauli basis $P$ and let $\Pi_{h}^{P}$ be the projector onto the basis state $\ket{h}_P$ of $\mathcal H_{\mathrm{hid}}$.

For any non-zero positive semidefinite operator $X$ on $\mathcal H_{\mathrm{vis}} \otimes \mathcal H_{\mathrm{hid}}$, define the completely positive trace-normalizing map
\[
\mathcal{G}_\beta(X) := e^{-(\beta/2) H}\,X\,e^{-(\beta/2) H},
\quad
\Gamma_\beta(X) := \frac{\mathcal{G}_\beta(X)}{\Tr[\mathcal{G}_\beta(X)]}.
\]
Then the conditional distributions of the sqRBM Gibbs state at inverse temperature $\beta$ are given by
\begin{align}
  p(h^P \mid v)
  &= \Tr\bigl[(\Iv \otimes \Pi_h^P)\,\Gamma_\beta \bigl(\Pi_v^Z \otimes 2^{-m}\Ih\bigr)\bigr], \label{eq:cond-given-v}\\
  p(v \mid h^P)
  &= \Tr\bigl[(\Pi_v^Z \otimes \Ih)\,\Gamma_\beta \bigl(2^{-n}\Iv \otimes \Pi_h^P\bigr)\bigr]. \label{eq:cond-given-h}
\end{align}
\end{theorem}

\begin{figure}[!t]
\centering
\begin{minipage}[]{\linewidth}
\subfloat[][Circuit for sampling $p(h^P\,\mid\,v)$]{
\begin{quantikz}
\lstick{$\ket{v}\!\bra{v}_\mathrm{Z}$}        & \qwbundle{n} & \gate[2]{\Gamma_\beta} &                          & \\
\lstick{$2^{-m} \Ih$} & \qwbundle{m} &                  & \meter{P-\mathrm{basis}} & \setwiretype{c} \rstick{$h^{P}$}\\
\end{quantikz}}
\label{fig:p(h|v)}
\end{minipage}

\begin{minipage}[]{\linewidth}
\subfloat[][Circuit for sampling $p(v\,\mid\,h^P)$]{
\begin{quantikz}
\lstick{$2^{-n} \Iv$} & \qwbundle{n} & \gate[2]{\Gamma_\beta} & \meter{\mathrm{Z}-\mathrm{basis}} & \setwiretype{c} \rstick{$v$} \\
\lstick{$\ket{h}\!\bra{h}_P$}    & \qwbundle{m} &                  &                          &  \\
\end{quantikz}}
\label{fig:p(v|h)}
\end{minipage}%
\caption{\textbf{Quantum circuits for conditional sampling in sqRBMs.} (a) Circuit realizing sampling from the conditional distribution  \(p(h^P\,\mid\,v)\). The visible subsystem is initialized in the computational basis state \(\ket{v}\), tensored with a maximally mixed hidden register, and evolved under $\Gamma_\beta$, which is described by \(\mathrm{exp}(-(\beta/2)H)\). (b) Circuit realizing sampling from the conditional distribution \(p(v\,\mid\,h^P)\), where the hidden subsystem is prepared in the Pauli basis state \(\ket{h}_P\) and the same evolution is applied. In both cases, measurement of the corresponding subsystem in its preparation basis yields statistics consistent with the conditional probabilities described in Theorem~\ref{theorem:cond-probs}.}
\label{fig:conditional-sampling-circuits}
\end{figure}
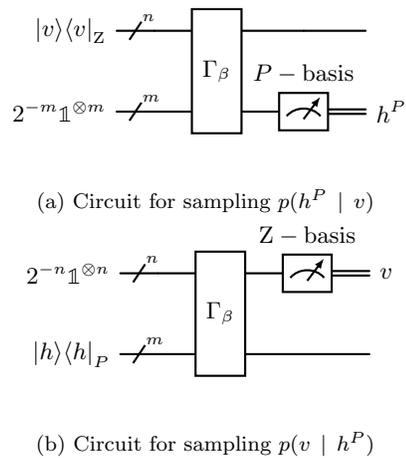

These conditional probabilities admit a direct quantum circuit realization. To obtain $p(h^P \mid v)$, one prepares the visible subsystem in the computational basis state $\ket{v}$, tensors it with a maximally mixed hidden register, applies the imaginary-time evolution operator $\mathrm{exp}(-(\beta/2)H)$, and measures the hidden subsystem in basis $P$. To obtain $p(v \mid h^P)$, one instead prepares the hidden subsystem in $\ket{h}_P$, tensors it with a maximally mixed visible register, applies the same operation, and measures the visible subsystem in the computational ($\Z$) basis. The resulting measurement statistics reproduce the conditional distributions of the sqRBM. The proof is provided in Appendix~\ref{app:conditional-proofs}, and the corresponding circuits are shown in Fig.~\ref{fig:conditional-sampling-circuits}.

\subsection{Generalized contrastive divergence}

Having established in Theorem~\ref{theorem:cond-probs} the explicit form of the conditional probabilities in sqRBMs, we can now formulate a generalized contrastive divergence (CD-$k$) procedure for training. The hidden operator pool $\mathcal{W}_h=\{\X,\Y,\Z\}$ can be partitioned into commuting subsets (i.e., $\vert\Wh\vert = 3$). This allows us to select a basis $P \in \mathcal{W}_h$, perform CD updates in that basis, and then repeat the process sequentially across all subsets to accumulate the full gradient contributions.

Each CD iteration begins with a visible configuration $v^{\mathrm{data}}$ drawn from the dataset. The positive phase sample of the hidden layer is obtained as $h^{(0), P} \sim p(h^P\,\mid\,v^{\mathrm{data}})$. A Gibbs chain of length $k$ is then run to generate negative phase samples: 
\begin{equation*}
v^{(s)} \sim p(v\,\mid\,h^{(s-1),\,P}), \quad h^{(s),\,P} \sim p(h^P\,\mid\,v^{(s)})
\end{equation*}
for $s=1, \dots, k$. As discussed in Section~\ref{sec:background}, in most applications a small number of Gibbs steps, often $k~=~1$, provides satisfactory performance~\cite{carreira-perpinan_contrastive_2005, sutskever_convergence_2010}. The final bitstrings $v^{(k)}$ and $h^{(k),\,P}$ serve as reconstructions samples.  

Then, the updates for the model parameters are computed from the four bitstrings $\{v^{\mathrm{data}}, h^{(0), P}, v^{(k)}, h^{(k),P}\}$:
\begin{align}
    &\Delta \boldsymbol{w}^{Z,P} = v^{\mathrm{data}} \cdot (h^{(0),P})^\top \;-\; v^{(k)} \cdot (h^{(k),P})^\top, \label{eq:genCD-w} \\
    &\Delta \boldsymbol{a} = v^{\mathrm{data}} - v^{(k)}, \label{eq:genCD-a} \\
    &\Delta \boldsymbol{b}^P = h^{(0),P} - h^{(k),P}. \label{eq:genCD-b}
\end{align}

Equations~\eqref{eq:genCD-w}–\eqref{eq:genCD-b} show that, for each operator basis $P$, a gradient update depends on conditional samples. Naively, this suggests that $|\Wh| \cdot (2k+1)$ conditional evaluations are needed to run a CD-$k$ update. A key simplification of sqRBMs is that their visible subspace is commuting. As a consequence, all hidden conditionals $p(h^P \mid v)$ can be sampled efficiently and in parallel on a classical computer, just as in standard RBMs. The only conditional that requires quantum resources is $p(v\,\mid\,h^P)$, which is used in the Gibbs chain and can be implemented via the circuits of Fig.~\ref{fig:conditional-sampling-circuits}. This means that the actual quantum cost of one CD-$k$ update reduces to $|\Wh| \cdot k$ circuit executions, each sampled once. The full procedure is summarized in Appendix~\ref{app:gcd}.

By analogy with deep neural network training, we refer to a single evaluation of the conditional distribution $p(v \mid h^P)$ through circuit execution and measurement as a \emph{forward pass}, since it represents the cost of generating samples from the Boltzmann machine’s visible distribution $p(v)$. For fixed $k$ and $|\mathcal{W}_h| = 3$, the asymptotic quantum cost of gradient computation is therefore $\mathcal{O}(1)$ forward passes per update.

\subsection{Numerical demonstrations}
\label{sec:numerics}

\begin{figure}[!t]
    \centering
    \includegraphics[width=\linewidth]{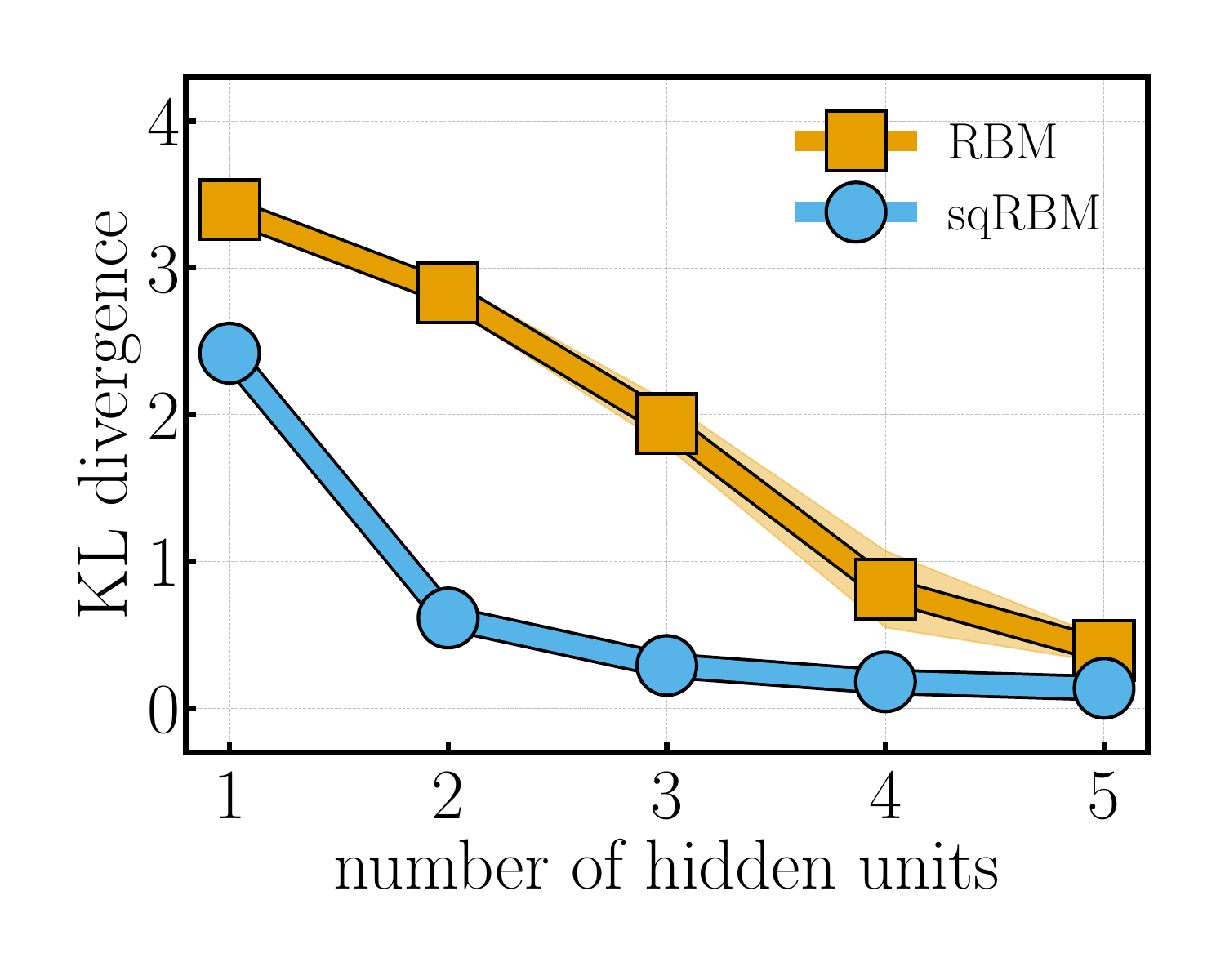}
    \caption{\textbf{KL divergence comparison between RBM and sqRBM trained with contrastive divergence.} Orange squares and blue circles denote RBM and sqRBM results on the bars and stripes (BAS) dataset, with shaded areas indicating one standard deviation over ten random initializations. The sqRBM reaches lower KL divergence with fewer hidden units.}
    \label{fig:rbm-sqrbm}
\end{figure}

The main motivation for studying sqRBMs lies in their ability to represent target probability distributions with fewer hidden units than classical RBMs. Ref.~\cite{demidik2025expressive} demonstrated this both analytically, through expressivity arguments, and empirically, via training based on minimizing the negative log-likelihood, as discussed in Section~\ref{sec:quantum-training}. In this work, we adopt sqRBMs as a representative model to showcase and benchmark generalized CD.

In the following, we use our proposed training algorithm to validate this result numerically. We train RBM and sqRBM models on the bars and stripes (BAS) dataset (details in Appendix~\ref{app:datasets}) with the number of hidden units ranging from one to five. Each configuration is initialized randomly ten times. The resulting KL divergences after training are shown in Fig.~\ref{fig:rbm-sqrbm}, and additional training details are provided in Appendix~\ref{app:trainig-details}.  

\begin{figure}[!t]
    \centering
    \includegraphics[width=\linewidth]{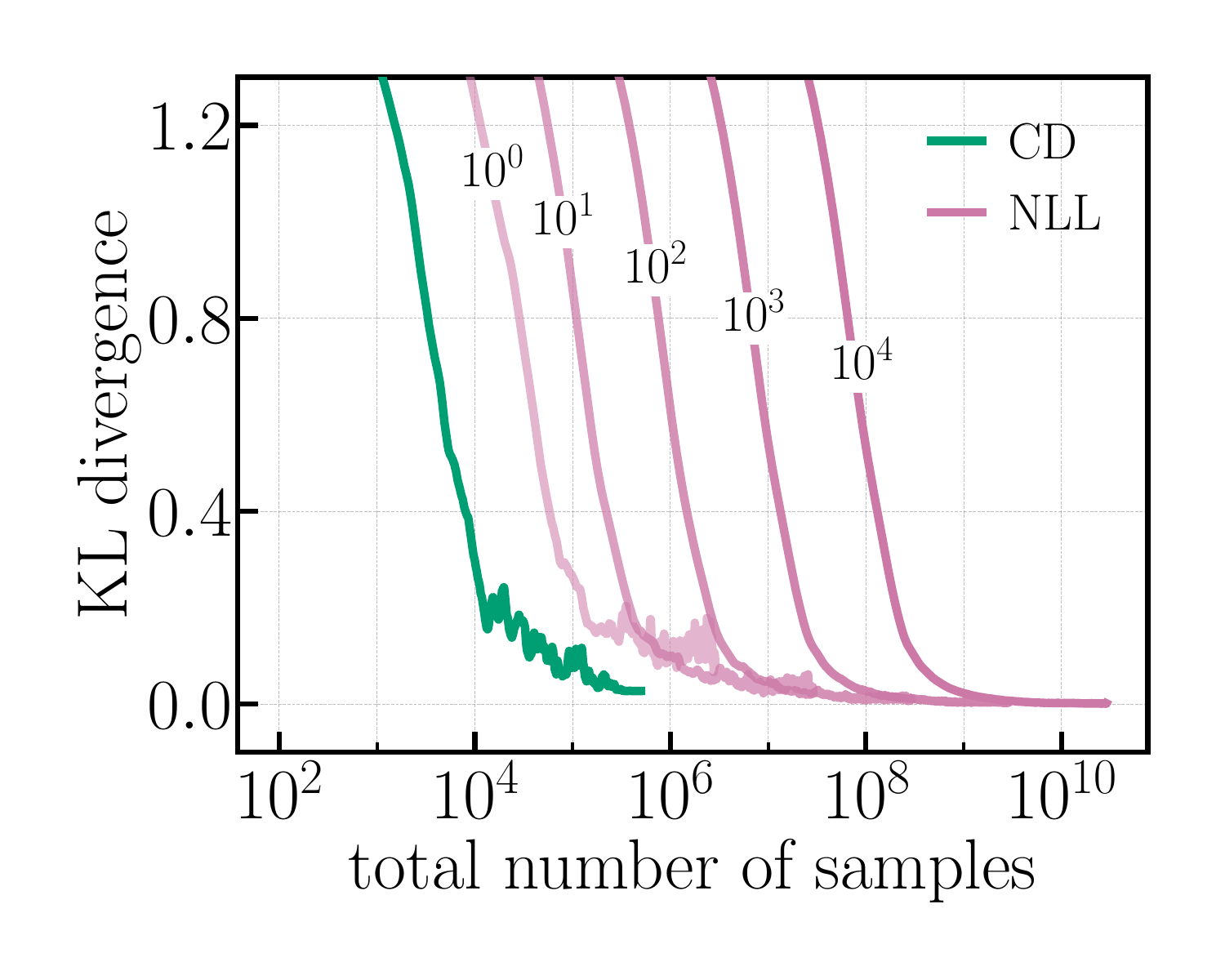}
    \caption{\textbf{Training curves comparing generalized contrastive divergence (CD) and negative log-likeligood (NLL) minimization.} The $\sqRBM_{9,3}$ model is trained on a Gaussian target distribution. The green line represents the proposed CD method, while pink lines of decreasing transparency correspond to NLL training (cf. Definition~\ref{def:sqbm-grads}) with 1 to $10^4$ shots per expectation value evaluation. The x-axis (log scale) indicates the total number of samples used to train models. The CD method achieves comparably low KL divergence during training while requiring significantly fewer samples.\looseness-1}
    \label{fig:cd-exact}
\end{figure}

In Fig.~\ref{fig:rbm-sqrbm}, we observe that the sqRBM exhibits a distinct scaling behavior compared to the classical RBM, reaching lower KL divergence values with fewer hidden units. This confirms that our training procedure functions reliably and is consistent with the findings of Ref.~\cite{demidik2025expressive}.

After verifying that our training algorithm performs as expected, we next compare its computational cost to the training based on negative log-likelihood (NLL) minimization. For this comparison, we use the $\mathrm{sqRBM}_{9,3}$ architecture and a Gaussian target distribution. In the NLL-based training, we assign a fixed shot budget for each expectation value evaluation, starting from one shot and increasing in powers of ten up to $10^4$.  We then compare the resulting performance to that of our generalized CD training under otherwise identical conditions. Technical details of this setup are provided in Appendix~\ref{app:trainig-details}.  

As shown in Fig.~\ref{fig:cd-exact}, the generalized CD approach trains the model to comparably low KL divergence values while requiring at least an order of magnitude fewer hardware samples. Although this difference is already visible at a small problem size, the gap is expected to grow with input size. As discussed in Section~\ref{sec:quantum-training}, the cost of NLL-based updates typically scales as $\mathcal{O}(\mathrm{poly}(n))$ with input size, since the gradient in Eq.~\eqref{eq:sqRBM-grad} requires computing expectation values over all visible configurations with non-zero support, whereas generalized contrastive divergence scales as $\mathcal{O}(1)$. It is worth noting that NLL minimization remains a principled procedure in the limit of infinite sampling, as it directly minimizes the KL divergence, while the generalized CD provides a stochastic approximation that achieves similar accuracy at substantially lower sampling cost.

\section{Discussion}\label{sec:discussion}

We introduce the generalized contrastive divergence training procedure, which exhibits $\mathcal{O}(1)$ scaling in the cost of a forward pass, representing a substantial improvement over previous QBM training methods~\cite{amin_quantum_2018, kappen_learning_2020, zoufal_variational_2021, patel2024natural, kimura2025em}. This scaling is analogous to the constant-cost parameter updates achieved by classical backpropagation, where gradients are obtained locally rather than through global evaluations~\cite{10.5555/65669.104451}. Such behavior has long been sought in quantum machine learning, since most existing algorithms exhibit linear scaling in number of parameters due to the parameter shift rule~\cite{abbas_quantum_2023, bowles_backpropagation_2025}. Our results therefore indicate that efficient gradient propagation can be realized within a fully quantum setting.

An equally important advantage of the generalized CD is that it eliminates the need for explicit access to the full model probability distribution. Earlier QBM training methods, based on gradient formulations such as Definition~\ref{def:sqbm-grads}, require evaluating expectation values over the complete distribution. This assumption makes them unsuitable for realistic generative modeling tasks, where the underlying distribution is unknown and only data samples are available. In contrast, the present approach enables training directly from samples, thereby allowing QBMs to operate as genuine generative models.

While our analysis focuses on the sqRBM architecture, the underlying principles are not specific to this model. The proof of Theorem~\ref{theorem:cond-probs} relies solely on the assumption that the model Hamiltonian is diagonal in the visible subspace. Consequently, the same training strategy can, in principle, be extended to architectures with more general connectivity and to deep QBMs, which are expected to exhibit richer representational behavior and offer greater potential for pattern recognition tasks~\cite{salakhutdinov_efficient_2012}. In the classical machine learning literature, it has been shown that CD performs reliably when exact samples can be drawn from the conditional distributions~\cite{Welling2002ANL}. This requirement is easier to satisfy in models with bipartite connectivity, where units of the same type are conditionally independent. Access to quantum hardware, however, may enable approximate yet sufficiently accurate sampling for more general connectivity architectures, thereby opening new directions for efficient training of Boltzmann machine variants.

Despite these advantages, the generalized CD algorithm remains a heuristic method without theoretical convergence guarantees. It has been shown that the parameter updates produced by classical CD-based training do not follow the gradient of any well-defined objective function~\cite{sutskever_convergence_2010}. Moreover, CD provides a biased estimate of the true likelihood gradient~\cite{bm_cd}. In practice, however, this bias is often mitigated by the method’s substantially lower variance and computational cost compared to full Gibbs sampling~\cite{bm_cd}. When sufficient sampling resources are available, likelihood-based methods such as NLL minimization provide a more principled optimization objective. A practical strategy may therefore combine both approaches by using generalized CD to obtain an approximate model of the target distribution, followed by fine-tuning through NLL minimization to achieve higher precision.

Furthermore, implementing the imaginary-time evolution underlying the Gibbs operator $\mathrm{exp}(-H)$ remains a major challenge on current quantum hardware. Developing efficient approximations of this evolution and compiling them into shallow, noise-resilient circuits is an active area of research~\cite{PhysRevLett.103.220502, temme_quantum_2011, 10.5555/3179483.3179486, holmesQuantumAlgorithmsFluctuation2022,wocjanSzegedyWalkUnitaries2021,rallThermalStatePreparation2023,zhangDissipativeQuantumGibbs2023,chenQuantumThermalState2023, chenEfficientExactNoncommutative2023,dingSimulatingOpenQuantum2024,gilyenQuantumGeneralizationsGlauber2024,dingEfficientQuantumGibbs2024,chenRandomizedMethodSimulating2024, hartung2025convergenceefficiencyproofquantum, brunnerLindbladEngineeringQuantum2025, chen2025}. Progress in this direction will directly benefit the practical realization of QBMs and, more broadly, the scalability of quantum generative models.

Finally, it is worth emphasizing that the generalized CD framework is not limited to Boltzmann machines. Extensions of classical CD have been successfully applied to a wide range of energy-based models (EBMs), including deep EBMs~\cite{luo_training_2023, du_improved_2021, yu_training_2020, 10.5555/3104482.3104621, xie_theory_2016, du_implicit_2020}. Extending this principle to quantum models would establish a foundation for sample-based training of quantum neural networks, providing a unifying perspective on gradient estimation across classical and quantum EBMs.

\begin{acknowledgments}
Authors thank Enrico Rinaldi, Matthias Rosenkranz and Marcello Benedetti for fruitful discussions. This work is supported with funds from the Ministry of Science, Research and Culture of the State of Brandenburg within the Centre for Quantum Technologies and Applications (CQTA). This work is funded within the framework of QUEST by the European Union’s Horizon Europe Framework Programme (HORIZON) under the ERA Chair scheme with grant agreement No.\ 101087126. C.T. and M.G. are supported by CERN through the CERN Quantum Technology Initiative. 
\end{acknowledgments}

\bibliography{main}

\newpage
\onecolumngrid
\newpage
\appendix

\newpage
\clearpage
\tableofcontents
\clearpage

\newpage
\section{Constrastive divergence}

\subsection{Training with contrastive divergence}\label{app:cd}

Training Boltzmann machines requires computing expectation values with respect to both the data and the model distributions. The latter is computationally intractable due to the exponential cost of obtaining exact samples from the Gibbs distribution. Consequently, approximate training techniques such as contrastive divergence (CD) are employed to mitigate this cost.

The idea behind CD is to approximate the gradient of the log-likelihood by contrasting statistics computed from the data distribution with those obtained from short Markov chains initialized at data samples. Instead of running these chains until equilibrium, CD-$k$ performs only $k$ alternating Gibbs updates, yielding a biased yet computationally efficient estimator of the likelihood gradient. In practice, this approximation performs remarkably well even for small values of $k$, typically $k=1$~\cite{carreira-perpinan_contrastive_2005, sutskever_convergence_2010}.

CD is primarily applied to the training of restricted Boltzmann machines (RBM), whose bipartite architecture allows the conditional distributions of visible and hidden units to factorize. This factorization enables efficient and parallel Gibbs updates across units:
\begin{align}
 p(h \mid v) &= \prod_{j=1}^m p(h_j \mid v), 
 &\quad p(h_j=1 \mid v) &= \sigma\!\left(b_j + \sum_{i=1}^n w_{ij} v_i\right), \\
 p(v \mid h) &= \prod_{i=1}^n p(v_i \mid h), 
 &\quad p(v_i=1 \mid h) &= \sigma\!\left(a_i + \sum_{j=1}^m w_{ij} h_j\right),
\end{align}
where $\sigma(x) = 1/(1+e^{-x})$ is the logistic sigmoid. Exact sampling from these conditionals is essential for CD training~\cite{Welling2002ANL}.

For completeness, we provide in Algorithm~1 (Fig.~\ref{alg:bm-cdk}) a pseudocode description of CD-$k$ training for RBMs. The update rules are obtained from the difference between the positive phase (data-driven samples) and the negative phase (reconstructions after $k$ Gibbs steps).

\begin{figure}[!h]
\centering
\begin{minipage}{\textwidth}
\hrule
\vspace{0.5em}
\textbf{Algorithm 1:} Contrastive Divergence with $k$ Gibbs steps (CD-$k$) for RBM training
\vspace{0.5em}
\hrule
\vspace{0.5em}
\ttfamily
\begin{tabbing}
\hspace{2em} \= \hspace{3em} \= \hspace{3em} \= \hspace{3em} \= \hspace{3em} \= \kill
1:\> \textbf{Input:} dataset $\mathcal{X}=\{v^{(\mu)}\}_{\mu=1}^M$, learning rate $\alpha$, number of iterations $T$, Gibbs steps $k$ \\
2:\> \textbf{Initialize} parameters $\boldsymbol{a}, \boldsymbol{b}, \boldsymbol{w}$ \\
3:\> \textbf{for} $\tau = 1$ to $T$ \textbf{do} \\
4:\> \> Sample data vector $v^{\mathrm{data}} \sim \mathcal{X}$ \\
5:\> \> \textbf{Positive phase:} \\
6:\> \> \> Sample hidden state $h^{(0)} \sim p(h\mid v^{\mathrm{data}})$ \\
7:\> \> \textbf{Negative phase:} \\
8:\> \> \> Initialize $(v^{(0)},h^{(0)}) = (v^{\mathrm{data}},h^{(0)})$ \\
9:\> \> \> \textbf{for} $s = 1$ to $k$ \textbf{do} \\
10:\> \> \> \> Sample $v^{(s)} \sim p(v \mid h^{(s-1)})$ \\
11:\> \> \> \> Sample $h^{(s)} \sim p(h \mid v^{(s)})$ \\
12:\> \> \> \textbf{end for} \\
13:\> \> \textbf{Parameter updates:} \\
14:\> \> \> $\Delta \boldsymbol{a} = v^{\mathrm{data}} - v^{(k)}$ \\
15:\> \> \> $\Delta \boldsymbol{b} = h^{(0)} - h^{(k)}$ \\
16:\> \> \> $\Delta \boldsymbol{w} = v^{\mathrm{data}} \cdot (h^{(0)})^\top- v^{(k)} \cdot (h^{(k)})^\top$ \\
17:\> \> Update parameters: \\
18:\> \> \> $\boldsymbol{a} \gets \boldsymbol{a} - \alpha \Delta \boldsymbol{a}$ \\
19:\> \> \> $\boldsymbol{b} \gets \boldsymbol{b} - \alpha \Delta \boldsymbol{b}$ \\
20:\> \> \> $\boldsymbol{w} \gets \boldsymbol{w} - \alpha \Delta \boldsymbol{w}$ \\
21:\> \textbf{end for} \\
22:\> \textbf{Return:} trained parameters $\boldsymbol{a}, \boldsymbol{b}, \boldsymbol{w}$
\end{tabbing}
\normalfont
\vspace{0.5em}
\hrule
\caption{Pseudocode to train RBM using contrastive divergence (CD-$k$). Parameter updates are obtained from the difference between the positive phase, computed from data samples, and the negative phase, approximated via a $k$-step Gibbs chain.}
\label{alg:bm-cdk}
\end{minipage}
\end{figure}

\subsection{Training with generalized contrastive divergence}\label{app:gcd}

For semi-quantum RBMs (sqRBMs), the training procedure must be adapted to account for the larger operator pool. The operator set $\Wh = \{\X,\Y,\Z\}$ can be separated into three mutually commuting subsets. This enables a generalized version of CD: one chooses a basis from these subsets, performs CD in that basis, and repeats sequentially to accumulate gradient contributions from all operators.

A consequence of Theorem~\ref{theorem:cond-probs} is that the conditional probabilities acquire a natural quantum circuit description. In the case of sqRBMs, however, the visible subspace remains commuting, which implies that sampling $p(h^P \mid v)$ can be carried out classically in the same efficient manner as in standard RBMs. Only the conditional distribution $p(v \mid h^P)$ requires quantum sampling, which can be implemented with the proposed circuit construction. The full procedure is summarized in Algorithm~2 (Figure~\ref{alg:gen-cdk}).

It is worth noting that Theorem~\ref{theorem:cond-probs} does not rely on an assumption of RBM-like bipartite connectivity. Consequently, the circuit-based conditionals derived there can be used to extend Algorithm~2 to more general semi-quantum Boltzmann machine architectures, including deep or fully connected variants. The only requirement for this generalization is that the visible subspace remains commuting.

\begin{figure}[!h]
\centering
\begin{minipage}{\textwidth}
\hrule
\vspace{0.5em}
\textbf{Algorithm 2:} Generalized Contrastive Divergence with $k$ Gibbs steps (CD-$k$) for sqRBM training
\vspace{0.5em}
\hrule
\vspace{0.5em}
\ttfamily
\begin{tabbing}
\hspace{2em} \= \hspace{3em} \= \hspace{3em} \= \hspace{3em} \= \hspace{3em} \= \kill
1:\> \textbf{Input:} dataset $\mathcal{X}=\{v^{(\mu)}\}_{\mu=1}^M$, learning rate $\alpha$, number of iterations $T$, Gibbs steps $k$, operator pool $\mathcal{W}$ \\
2:\> \textbf{Initialize} parameters $\boldsymbol{a}, \boldsymbol{b}, \boldsymbol{w}$ \\
3:\> \textbf{for} $\tau = 1$ to $T$ \textbf{do} \\
4:\> \> Sample data vector $v^{\mathrm{data}} \sim \mathcal{X}$ \\
5:\> \> \textbf{for each} $P \in \mathcal{W}$ \textbf{do} \\
6:\> \> \> \textbf{Positive phase:} sample $h^{(0),P} \sim p(h^P \mid v^{\mathrm{data}})$ \\
7:\> \> \> \textbf{Negative phase (CD-$k$):} set $(v^{(0)},h^{(0),P}) \gets (v^{\mathrm{data}},h^{(0),P})$ \\
8:\> \> \> \textbf{for} $s = 1$ to $k$ \textbf{do} \\
9:\> \> \> \> Sample $v^{(s)} \sim p(v\mid h^{(s-1),P})$ \\
10:\> \> \> \> Sample $h^{(s),P} \sim p(h^P \mid v^{(s)}$ \\
11:\> \> \> \textbf{end for} \\
12:\> \> \> \textbf{Parameter updates:} \\
13:\> \> \> \> $\Delta \boldsymbol{a} \;=\; v^{\mathrm{data}} - v^{(k)}$ \\
14:\> \> \> \> $\Delta \boldsymbol{b}^{P} \;=\; h^{(0),P} - h^{(k),P}$ \\
15:\> \> \> \> $\Delta \boldsymbol{w}^{Z,P} \;=\; v^{\mathrm{data}} \cdot (h^{(0),P})^\top - v^{(k)} \cdot (h^{(k),P})^\top$ \\
16:\> \> \> \textbf{Update parameters:} \\
17:\> \> \> $\boldsymbol{a} \gets \boldsymbol{a} - \alpha\, \Delta \boldsymbol{a}$ \\
18:\> \> \> $\boldsymbol{b} \gets \boldsymbol{b} - \alpha\, \Delta \boldsymbol{b}^P$ \\
19:\> \> \> $\boldsymbol{w} \gets \boldsymbol{w} - \alpha\, \Delta \boldsymbol{w}^{Z,P}$ \\
20:\> \> \textbf{end for} \\
21:\> \textbf{end for} \\
22:\> \textbf{Return:} trained parameters $\boldsymbol{a}, \boldsymbol{b}, \boldsymbol{w}$
\end{tabbing}
\normalfont
\vspace{0.5em}
\hrule
\caption{Pseudocode to train sqRBM using the generalized contrastive divergence (CD-$k$).}
\label{alg:gen-cdk}
\end{minipage}
\end{figure}

\newpage
\section{Derivations and proofs}

\subsection{Useful definitions}

We define the following useful identities for marginal and joint distributions in the Hamiltonian formalism:
\begin{align}
    &p(v)   = \Tr[(\Pi_v^Z \otimes \Ih) \rho] \\
    &p(h^P) = \Tr[(\Iv \otimes \Pi_h^P) \rho] \\
    &p(v, h^P) = \Tr[(\Pi_v^Z \otimes \Pi_h^P) \rho] 
\end{align}

\subsection{Conditional probabilities of semi-quantum RBMs}\label{app:conditional-proofs}

\begin{proof}
We prove the first identity in Eq.~\eqref{eq:cond-given-v}; the second follows analogously by exchanging the visible and hidden subsystems.

For the Gibbs state
\(
\rho = e^{-\beta H} / \Tr[e^{-\beta H}],
\)
the joint probability of observing outcomes $v$ and $h^P$ in the visible and hidden subsystems is
\begin{equation}
    p(v, h^P) = \Tr[(\Pi_v^Z \otimes \Pi_h^P)\,\rho].
\end{equation}
The marginal probability of $v$ is obtained by summing over the hidden outcomes in the computational basis,
\begin{equation}
    p(v) = \sum_{h} p(v, h^Z)
          = \Tr[(\Pi_v^Z \otimes \Ih)\,\rho].
\end{equation}

By Bayes’ rule, the conditional probability is therefore
\begin{equation}\label{eq:cond-start}
    p(h^P \mid v)
    = \frac{p(v, h^P)}{p(v)}
    = \frac{\Tr[(\Pi_v^Z \otimes \Pi_h^P)e^{-\beta H}]}
           {\Tr[(\Pi_v^Z \otimes \Ih)e^{-\beta H}]}.
\end{equation}

Let us insert $e^{-\beta H} = e^{-(\beta/2)H} e^{-(\beta/2)H}$ and use the cyclicity of the trace to obtain
\begin{equation}
p(h^P \mid v)
= \frac{\Tr\!\left[(\Iv \otimes \Pi_h^P)
   e^{-(\beta/2)H} (\Pi_v^Z \otimes \Ih) e^{-(\beta/2)H}\right]}
   {\Tr\!\left[e^{-(\beta/2)H} (\Pi_v^Z \otimes \Ih) e^{-(\beta/2)H}\right]}.
\end{equation}

The visible projectors commute with the Hamiltonian of an sqRBM (cf. Definition~\ref{def:sqrbm}), $[\Pi_v^Z, H]=0$, allowing $\Pi_v^Z$ to be moved through the exponentials. Multiplying numerator and denominator by $2^{-m}$ and identifying
\[
\mathcal{G}_\beta(X) = e^{-(\beta/2)H} X e^{-(\beta/2)H},
\qquad
\Gamma_\beta(X) = \frac{\mathcal{G}_\beta(X)}{\Tr[\mathcal{G}_\beta(X)]},
\]
we obtain
\begin{equation}
p(h^P \mid v)
= \Tr\!\big[(\Iv \otimes \Pi_h^P)
   \Gamma_\beta(\Pi_v^Z \otimes 2^{-m}\Ih)\big],
\end{equation}
which reproduces Eq.~\eqref{eq:cond-given-v}.

The second identity, Eq.~\eqref{eq:cond-given-h}, follows analogously by interchanging $\Pi_v^Z$ and $\Pi_h^P$ and replacing $2^{-m}$ with $2^{-n}$.

Operationally, both conditionals correspond to imaginary-time evolution of reference states
\[
\rho_0(v) = \ket{v}\!\bra{v} \otimes 2^{-m}\Ih,
\qquad
\rho_0(h^P) = 2^{-n}\Iv \otimes \ket{h}\!\bra{h}_P,
\]
under $\mathcal{G}_\beta$. Measurements of the complementary subsystem in the respective basis then yield samples distributed according to
$p(h^P \mid v)$ and $p(v \mid h^P)$.
\end{proof}

\subsection{Closed form conditional probabilities of sqRBM models for classical simulation}

In this subsection, we provide supplementary derivations that were used to simulate conditional distributions in the sqRBM model. Owing to their structured form, sampling from sqRBMs can be performed more efficiently than full diagonalization of the corresponding Gibbs state. Note, however, that these are unnormalized distributions; consequently, the resulting probabilities do not in general factorize into a product form (although they may do so in specific cases discussed in the main text). Therefore, a quantum device is still required to perform sampling efficiently. The derivations presented here are intended for completeness and are best read in conjunction with Ref.~\cite{demidik2025expressive}, which our results extend. 

Let us define:
\begin{equation}
    \phi_j^{P}(v, h_j) = b_{j}^{P} (-1)^{h_j} + \sum_{i=1}^{n} (-1)^{v_i + h_j} w_{i,j}^{Z,P},
    \label{eq:hidden_state}
\end{equation}

\begin{equation}
    \Phi_j(v, h_j) = \left[\phi_j^{X}(v, h_j) \quad \phi_j^{Y}(v, h_j) \quad \phi_j^{Z}(v, h_j)    \right]
\end{equation}

\begin{equation}
    D_j^P(v, h_j^P) = \cosh(||\Phi_j(v, h_j^P)||_2)-\frac{\phi_j^P(v, h_j^P)}{||\Phi_j(v, h_j^P)||_2} 
    \sinh(||\Phi_j(v,h_j^P)||_2).    
\end{equation}

Then, the unnormalized joint probability distributions take the form:
\begin{equation}
    \tilde{p}(v \,\mid \, h^P) = \prod_{i=1}^{n} (-1)^{v_i} \prod_{j=1}^{m} D_j^P(v, h_j^P)\quad \text{and}\quad \tilde{p}(h^P\, \mid \,v) = \prod_{j=1}^{m} D_j^P(v, h_j^P).
\end{equation}

\newpage
\section{Details of numerical demonstrations}
\label{app:numerical-details}

\subsection{Visualization of datasets}
\label{app:datasets}

\begin{figure}[!h]
\centering
\begin{minipage}[b]{.49\linewidth}
\subfloat[][Bars and stripes (BAS) dataset]{
\includegraphics[width=\linewidth]{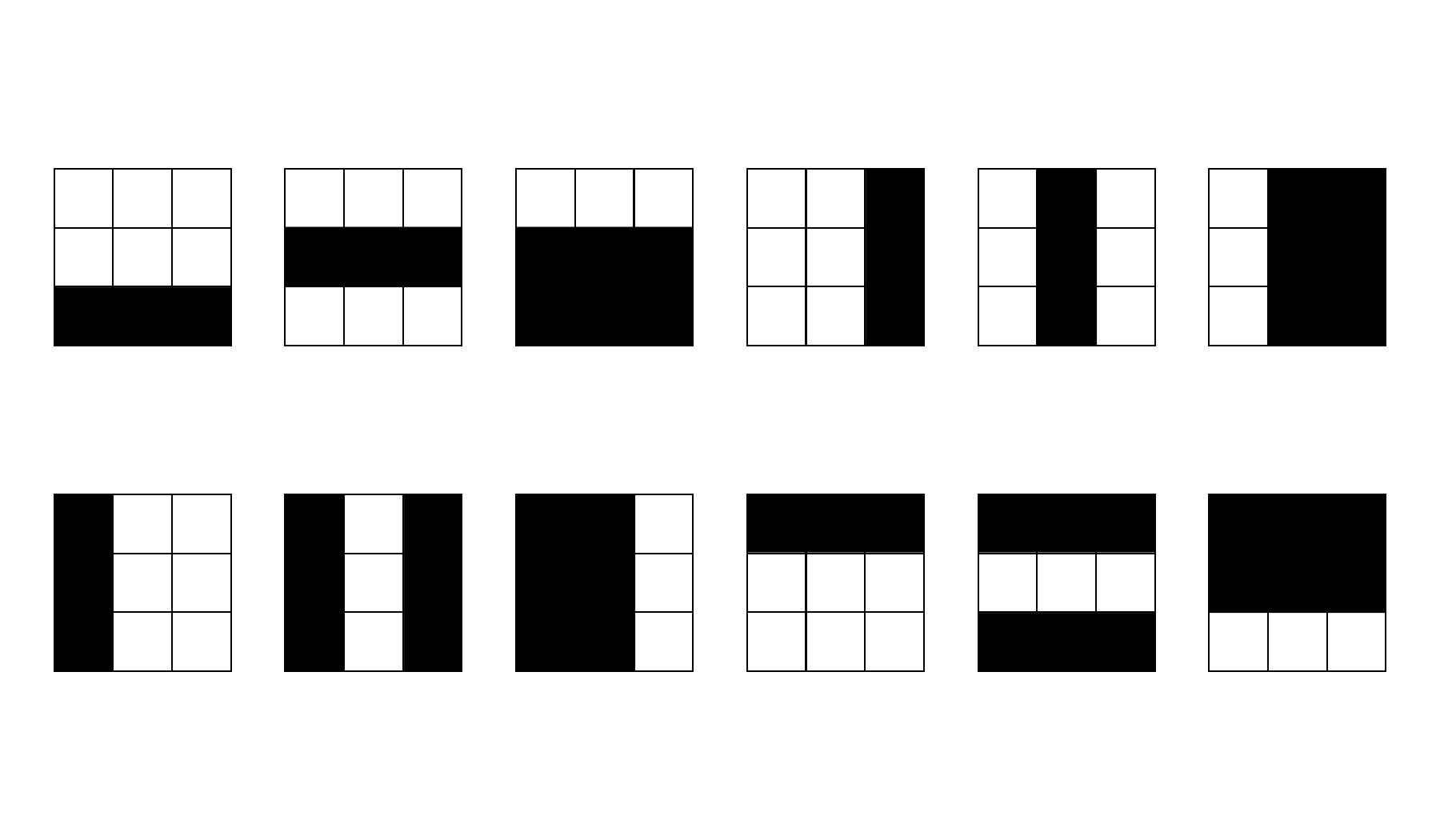}
\label{fig:bas-dataset}
}
\end{minipage}
\hfill
\begin{minipage}[b]{.49\linewidth}
\subfloat[][Gaussian dataset]{
\includegraphics[width=\linewidth]{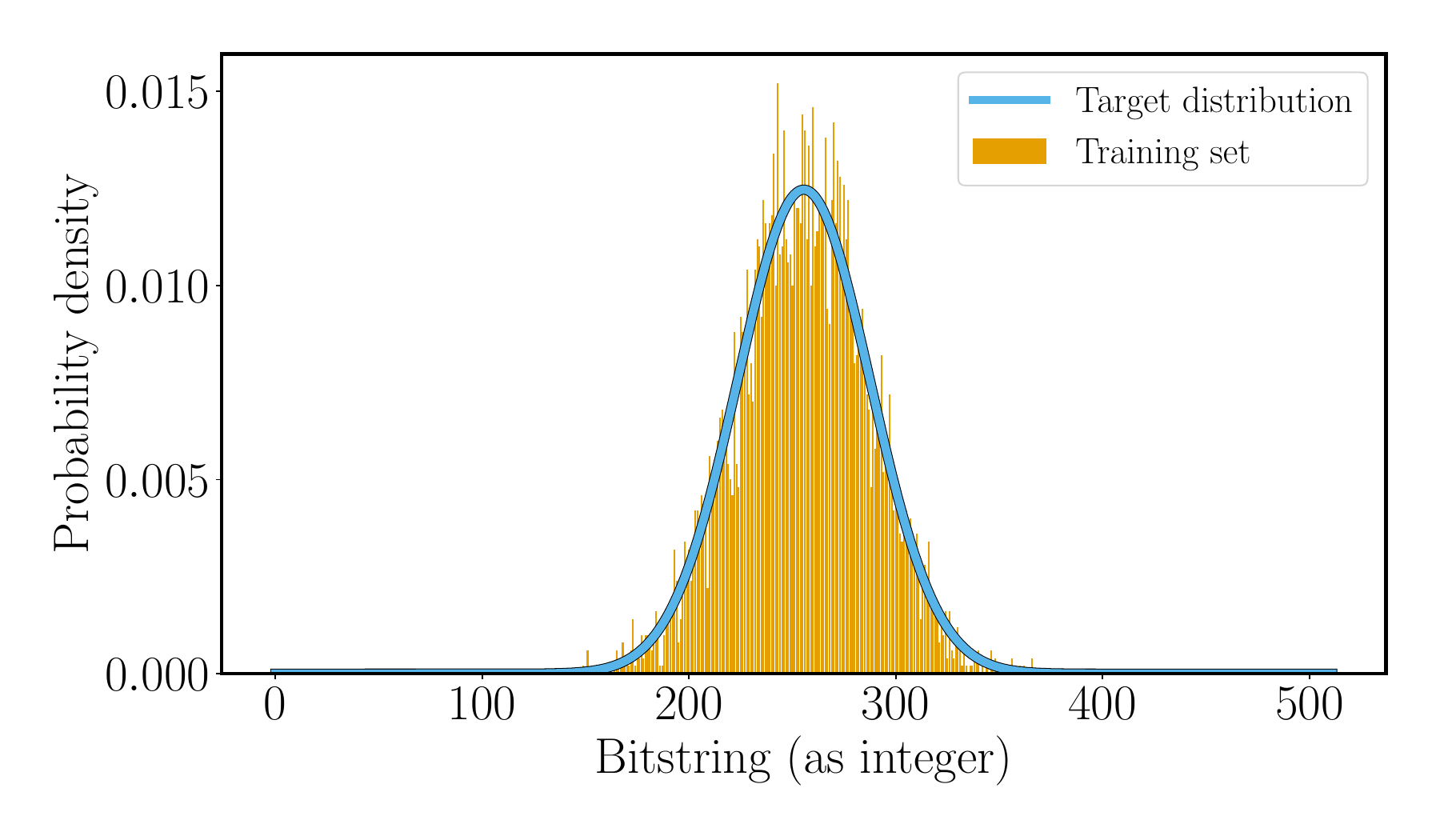}
\label{fig:}
}
\end{minipage}

\caption{\textbf{Datasets used for numerical demonstrations} (a) Bars and stripes (BAS) dataset defined on a $3 \times 3$ grid. The dataset consists of 12 unique entries. (b) Discretized Gaussian distribution $\mathcal{N}(\mu{=}255.5,\, \sigma{=}32)$ defined over 512 bins ($2^9$ bitstrings). The solid line shows the exact distribution, and the bars represent 10,000 samples drawn from it to form the training dataset.}
\label{fig:gaussian-dataset}
\end{figure}

\subsection{Training details}
\label{app:trainig-details}

All models were trained for \(5000\) iterations using the ADAM optimizer with parameters \(\epsilon=10^{-8}\), \(\beta_1=0.9\), and \(\beta_2=0.99\). The initial learning rate was set to \(\eta_0=0.01\) and exponentially decayed to a minimum of \(10^{-4}\) by step \(T=5000\). Gradients were clamped elementwise to the range \([-0.05, 0.05]\). Contrastive divergence training relied on \(k=10\) Gibbs steps and ten persistent Markov chains. These settings are representative of standard practices for Boltzmann machine models and sufficiently capture the qualitative training dynamics of interest, so no dedicated hyperparameter search was performed.

For the BAS dataset, weights were initialized from \(\mathcal{N}(0,\sigma)\) with variance \(\sigma^2 = 10/(n+m)\), hidden biases were set to zero, and visible biases were initialized such that the model’s single-bit marginals matched the empirical averages. The models were trained with full batch size (\(12\)) and a final learning rate of \(10^{-3}\) after decay. 

For the Gaussian dataset, weights were drawn from \(\mathcal{N}(0,\sigma)\) with \(\sigma^2 = 1/(n+m)\), and the same bias initialization scheme was applied. We used an \(\sqRBM_{9,3}\) architecture with $\Wh = \{\X,\Y,\Z\}$. To estimate the cost of evaluating the negative log-likelihood (NLL), we consider the data support to be \(\vert \mathrm{supp}(q)\vert=191\) out of \(2^9=512\) configurations corresponding to samples with reliable signal. Moreover, we assume that the Hamiltonian can be partitioned into three mutually commuting subsets so that all expectation values \(\langle H_i\rangle\) can be obtained using three measurement settings. We further assume that evaluating expectation values on the projected state (cf. Definition~\ref{def:sqbm-grads}) has the same cost as on the Gibbs state. Although slightly idealized, these assumptions make the NLL baseline comparison maximally favorable.

\end{document}